\documentclass[preprintnumbers,article,amsmath,amssymb,floatfix,10pt,prd,onecolumn,
superscriptaddress,nofootinbib]{revtex4}
\usepackage{bbm}
\usepackage{amsfonts}
\usepackage{mathrsfs}
\usepackage{latexsym}

\usepackage{epsfig}
\usepackage{epstopdf}
\usepackage{graphicx}
\usepackage{amssymb}
\usepackage{amsmath}
\usepackage{dcolumn}
\usepackage{bm}
\usepackage{color}
\usepackage{comment}
\usepackage{xcolor}
\begin{document}

\title{$f(\mathcal{G},\mathrm{\textit{T}})$ Gravity Bouncing Universe with Cosmological Parameters}

\author{Mushtaq Ahmad}
\email{mushtaq.sial@nu.edu.pk}\affiliation{National University of Computer and
Emerging Sciences, Islamabad,\\ Chiniot-Faisalabad Campus, Pakistan.}
\author{M. Farasat Shamir}
\email{farasat.shamir@nu.edu.pk; farasat.shamir@gmail.com}\affiliation{National University of Computer and
Emerging Sciences, Islamabad, \\ Lahore Campus, Pakistan.}
\author{G. Mustafa}
\email{gmustafa3828@gmail.com}\affiliation{Department of Mathematics, Shanghai University,
Shanghai, 200444, Shanghai, People's Republic of China}

\begin{abstract}
In recent few years, the Gauss-Bonnet $f(\mathcal{G},\mathrm{\textit{T}})$ theory of gravity has fascinated considerable researchers owing to its coupling of trace of the stress-energy tensor $T$ with the Gauss-Bonnet term $\mathcal{G}$. In this context, we focuss ourselves to study bouncing universe with in $f(\mathcal{G},\mathrm{\textit{T}})$  gravity background. Some important preliminaries  are presented along with the discussion of cosmological parameters to develop a minimal background about $f(\mathcal{G},\mathrm{\textit{T}})$  theory of gravity. The exact bouncing solutions with physical analysis are provided with the choice of two equation of state parameters. It is shown that the results do agree with the present values of deceleration, jerk and snap parameters. Moreover, it is concluded that the model parameters are quite important for the validity of conservation equation (as the matter coupled theories do not obey the usual conservation law).

\end{abstract}

\maketitle

\date{\today}

\section{Introduction}


The phenomenon of accelerating expansion of the Universe is now quite obvious to us all either in the early stage
of evolution also known as the inflation or the phase of the late Universe. Innumerable cosmological authentic observations provide the indication of the late time cosmic acceleration. Some of these observations include cosmic microwave background radiations (CMBR) \cite{Spergel1}-\cite{Spergel02},  high red-shift supernovae \cite{ReissExp1}, Planck data \cite{AdeExp6}, supernovae of type Ia \cite{PerlmutterExp2}-\cite{BennettExp3}, and baryon acoustic oscillations \cite{PercivalExp7} . The ultimate foundation of the recent cosmology is, no doubt, the theory of general relativity (GR) as it theoretically enlightens the large-scale structure of our Universe splendidly. So as to elucidate the epoch of late time cosmic acceleration, diverse theories have been structured in the earlier decades like the modification of the geometry in the Einstein's field equations (EFEs) (by modifying the energy momentum tensor). The addition of the matter term to the extremely negative pressure called as the dark energy (DE) included in EFEs successfully explains the mystery of the late-time accelerating epoch of the Universe as revealed by the observations. But, the mysterious nature of  DE is still not known to us as there does not exist reliable evidence of DE. Largely, it is believed that the DE is a homogenized form of some fluid that fills through the space contributing nearly two-third of the entire Universe. Nevertheless, this is a matter which demands a great discussion among the cosmologists on the role of DE.
As suggested by the  Big Bang, the Universe started from a singularity
occurring in the space-times which has some limitations e.g. flatness problem, horizon problem, transplanckian, original structure problem, entropy problem,  and the singularity problem. To settle these issues, an unexpected evolution of Universe was obvious after the big bang to establish its smooth and uniform structure. Alan Guth developed inflationary theory to explain diverse standard cosmological complications which emerged as  a successful theory to illustrate the several observational features of the expanding Universe. Though, the problem of initial singularity being a fundamental problem still remains unanswered in the recent era of cosmology.\par

As a test bed, cosmology plays a crucial role for the theories beyond the general model exhibited by the Big Bang theory, together with the theories of quantum gravity and string-inspired theories. The existing observational data vigorously favors the primordial inflation, nearing to the grand uniﬁed scale \cite{Spergel1}. Inflation has proven to be a dynamical explanation of the problems such as horizon, ﬂatness,  and homogeneity \cite{Linde2}. In a standard model, for instance, the one motivated by some scalar ﬁeld, it also gives metric fluctuations beyond the Hubble radii through a power spectrum of scale invariant \cite{Mukhanov3}. Regardless of the tremendous achievements of inflation, it unfortunately does not report the crucial features of the Big Bang cosmology. For any equation of state (EoS) favoring the strong energy condition (SEC) $-\rho/3<p$, the scale factor of the universe in a Friedmann Robertson Walker (FRW) metric vanishes at $t=0$, irrespective of the universe geometry (either ﬂat, open, or closed) and the energy density deviates. In fact, the entire curvature invariants, such as $R$, $\Box{R}$, etc. turn out to be singular. This is the because of the reason that it has been acknowledged as the Big Bang singularity problem.
Even though the inflation justifies $-\rho/3<p$, it does not relieve the Big Bang singularity problem, it pushes rather in time the singularity backwards. Several authors have considered this issue that either the inflation is past eternal or not \cite{Linde4}-\cite{Borde5}, and it is concluded that, it is not in the framework of Einstein gravity at least provided that the average accelerated rate of expansion formerly is bigger than zero, that is, $H_{av} > 0$ \cite{Borde6}. The ﬂuctuations are produced with the universe approaching the singularity, and the Hawking and Penrose standard singularity theorems hold, which certainly leads to an ultimate collapse in FRW universe given that the energy density remains positive \cite{Hawking7}. Many other attempts have been made to avoid the Big Bang singularity problem occurring as a result of anisotropic stresses, quantum cosmology, self-regenerating (during inﬂation ) universes, etc. \cite{Linde2}, nevertheless, none has provided any valid solution to the problems such as the space-like singularity, specifically with reference to the flat universe. Even the string-inspired theories have yet to address this problem comprehensively: Quite a lot of toy model structures have been presented, but most of them encounter some deviations, such as quantum instabilities, closed time-like curves, negative-tension branes,  presence of ghosts, singular bounce, etc.\par

One of the striking possible substitutes to this inflationary model was presented as the bouncing cosmology model to explain the expanding Universe which cracks the mid 1980's problem of initial singularity. This model says that the Universe might have emerged from some prior phase of contraction being capable of expanding without experiencing any singularity or it undergoes an ultimate bouncing process \cite{Shabani43}-\cite{Kuiroukidis44}.
Qiu et al \cite{aanu} presented nonsingular, homogeneous and isotropic bouncing solutions of the conformal Galileon model. In their study, they proved that such solutions necessarily began with a radiation-dominated contracting phase. They also predicted a blue tilted spectrum of gravitational waves stemming from quantum vacuum fluctuations in the contracting phase. In another work \cite{manu}, a noncanonical minimally coupled scalar field belonging to the class of theories with Galileon-like self-couplings is used to show a smooth evolution from contraction to expansion phase. Some necessary restrictions were formulated for Lagrangians needed to obtain a healthy bounce.
Recently, some exceptional research works have been accomplished by Cai et al. \cite{Cai46} and Brandenberger et al. \cite{Brandenberger50} explaining the nonsingular phenomenon of bouncing cosmology. In their work, they used cosmological models on the Galileon theory to study some important features of the bouncing cosmology.

For the description of early universe by a cosmological model, the matter bounce scenario is recognized \cite{Brandenberger1}.
In this situation, the universe is dominated by matter in the contraction stage, and a singularity-free bounce happens. Likewise, the density perturbations possessing the consistent spectrum with the observations may be formed \cite{Novello2}. Additionally, after the contracting stage, the theoretical BKL instability \cite{Belinsky3} occurs in a way that the universe will become anisotropic. The way of circumventing this instability \cite{Erickson4}and the problems of the bounce \cite{Xue5} in the Ekpyrotic phenomenon \cite{Khoury6} has been examined \cite{Cai7}-\cite{Cai8}. Furthermore, the density perturbations within the matter bounce scenario containing two scalar fields has been examined in a recent work \cite{Nojiri9}.
In different circumstances, a number of cosmological observations favor the ongoing accelerated cosmos expansion. To support this phenomenon in the isotropic and homogeneous universe, it is important to consider the presence of DE exhibiting negative pressure, or assume the gravity is modified on large scales \cite{Nojiri10}-\cite{Bamba11}. As for the latter approach is concerned, a number of modified gravity theories such as $f(R)$, $f(\mathcal{G})$, $f(R,T)$, $f(\mathcal{G}, T)$, and many others have been presented. The bouncing aspects have been explored in $f(R)$ gravity \cite{Olmo 12}-\cite{Bamba15}, string-associated theories of gravitational \cite{Biswas16}, non-local gravity \cite{Biswas17}. A connection between the bouncing conduct and the variances of the cosmic microwave background (CMB) radiation has been discussed as well in \cite{Liu18}.

Thus, motivated from above discussion, we are focussed to investigate $f(\mathcal{G},\mathrm{\textit{T}})$ gravity bouncing Universe in the light of important cosmological parameters. The paper is planned in the following way: Section $2$ provides some important preliminaries along with the discussion of cosmological parameters to develop a minimal background about $f(\mathcal{G},\mathrm{\textit{T}})$ theory of gravity. The exact bouncing solutions with physical discussion is added in third section. Last section summarize the results.

\section{$f(\mathcal{G},\mathrm{\textit{T}})$ Gravity: Some Important Preliminaries}

The Gauss-Bonnet $f(\mathcal{G},\mathrm{\textit{T}})$ theory of gravity in recent years  has fascinated several researchers owing to its coupling of trace of the stress-energy tensor $T$ with the GB term $\mathcal{G}$. The action of this modified gravity is  described by the following integral \cite{sharif.ayesha},
\begin{equation}\label{action}
\mathcal{A}= \frac{1}{2{\kappa}^{2}}\int d^{4}x
\sqrt{-g}[R+f(\mathcal{G},\mathrm{\textit{T}})]+\int
d^{4}x\sqrt{-g}\mathcal{L}_{M}.
\end{equation}
The $f(\mathcal{G},\mathrm{\textit{T}})$ theory of gravity has shown great competence in answering some critical problems emerging since the advent of GR. But, a frequently faced problem by such modified theories possessing high curvature derivatives is contained in the higher-order terms incorporating the stress-energy tensor. This causes the divergence from the standard equation providing energy-momentum conservation.
Consequently, a small worrying situation involves when the theory could be affected due to the existence of such divergences at the cosmological scales. However, the issue might be controlled by pushing it to some limitations on its divergence equation. Such implication is expected as well to support the recovery of the standard conservation equation. Currently, we are interested in working out the bouncing cosmology under the  $f(\mathcal{G},\mathrm{\textit{T}})$ gravity theory by employing the flat FLRW geometry.

\begin{equation}\label{4}
ds^{2}=dt^2-a^2(t)[dx^2+dy^2+dz^2],
\end{equation}
where $a$ being the cosmic scale factor. Now by varying the action ($\ref{action}$) about the metric tensor $g_{\xi\eta}$ provides the following field equation under the $f(\mathcal{G},\mathrm{\textit{T}})$ gravity \cite{sharif.ayesha}
\begin{eqnarray}\nonumber
&R_{\xi\eta}-\frac{1}{2}g_{\xi\eta}R+[2Rg_{\xi\eta}\nabla^{2}-2R\nabla_{\xi}\nabla_{\eta}-4g_{\xi\eta}R^{\mu\nu}\nabla_{\mu}\nabla_{\nu}-
4R_{\xi\eta}\nabla^{2}+4R^{\mu}_{\xi}\nabla_{\eta}\nabla_{\mu}+
4R^{\mu}_{\eta}\nabla_{\xi}\nabla_{\mu}+4R_{\xi\mu\eta\nu}\nabla^{\mu}\nabla^{\nu}]f_{\mathcal{G}}(\mathcal{G},\mathrm{\textit{T}})\\&-
\frac{1}{2}g_{\xi\eta}f(\mathcal{G},\mathrm{\textit{T}})+
[\mathrm{\textit{T}}_{\xi\eta}+\Theta_{\xi\eta}]f_{\mathrm{\textit{T}}}(\mathcal{G},\mathrm{\textit{T}})+
[2RR_{\xi\eta}-4R^{\mu}_{\xi}R_{\mu\eta}-4R_{\xi\mu\eta\nu}R^{\mu\nu}+2R^{\mu\nu\delta}_{\xi}R_{\eta\mu\nu\delta}]
f_{\mathcal{G}}(\mathcal{G},\mathrm{T})=-\kappa^{2}\mathrm{\textit{T}}_{\xi\eta},\label{4_eqn}
\end{eqnarray}
where all the symbols appearing above or after depict their standard meanings, and the subscripts $\textit{T}$ and $\mathcal{G}$ involved in the functions provide the corresponding partial derivatives. Also, $\Theta_{\xi\eta}= g^{\mu\nu}\frac{\delta
\mathrm{\textit{T}}_{\mu\nu}}{\delta g_{\xi\eta}}$. Now, when taken the trace of Eq.($\ref{4_eqn}$), reads
\begin{equation}\label{5_eqn}
R+\kappa^{2}\mathrm{\textit{T}}-(\mathrm{\textit{T}}+\Theta)f_{\mathrm{\textit{T}}}(\mathcal{G},\mathrm{T})+2f(\mathcal{G},\mathrm{T})
+2\mathcal{G}f_{\mathcal{G}}(\mathcal{G},\mathrm{T})-2R\nabla^{2}f_{\mathcal{G}}(\mathcal{G},\mathrm{T})
+4R^{\xi\eta}\nabla_{\xi}\nabla_{\eta}f_{\mathcal{G}}(\mathcal{G},\mathrm{T})=0.
\end{equation}
Interestingly, after placing $f(\mathcal{G},\mathrm{\textit{T}})=0$ into Eq.($\ref{5_eqn}$), the GR equations are retrieved as
\begin{equation}\label{05_eqn}
R+\kappa^2\mathrm{\textit{T}}=0.
\end{equation}
The fascinating feature of Eq.($\ref{5_eqn}$) is to exhibit the correspondence of the terms $\mathcal{G}$, $\mathrm{\textit{T}}$, and $R$, in diverse manner. However, as evident by the corresponding GR version in Eq.($\ref{05_eqn}$), $R$ and $\mathcal{G}$ are worked out mathematically. This undoubtedly advocates that the modified equations may adhere several solutions than those of the GR.
Following Eq.(\ref{4_eqn}) expresses the covariant divergence as
\begin{equation}\label{div}
\nabla^{\xi}T_{\xi\eta}=\frac{f_{\mathrm{\textit{T}}}(\mathcal{G},\mathrm{T})}
{\kappa^{2}-f_{\mathrm{\textit{T}}}(\mathcal{G},\mathrm{T})}\bigg[(\mathrm{\textit{T}}_{\xi\eta}+\Theta_{\xi\eta})
\nabla^{\xi}(\text{Log}f_{\mathrm{\textit{T}}}(\mathcal{G},\mathrm{T}))+
\nabla^{\xi}\Theta_{\xi\eta}-
\frac{g_{\xi\eta}}{2}\nabla^{\xi}\textit{T}\bigg].
\end{equation}
It is quite clear that the above equation does not vanish as it happens in case of the GR. Though, one may opt to indulge some conditions on the Eq.(\ref{div}) to tackle the discrepancy so as to retrieve the corresponding standard equation. The simplest option here is to assume the expression surrounded by the parenthesis equaling to zero. For our ongoing work, we invest our efforts in finding the cosmology in {$f(\mathcal{G},\mathrm{\textit{T}})$ theory by supposing that the universe with perfect matter distribution, that is
\begin{equation}\label{7}
\textit{T}_{\mu\nu}=(\rho + p)u_\mu u_\nu-pg_{\mu\nu},
\end{equation}
where the physical parameters $\rho$ and $p$ stand for the energy density and stress of the matter involved, respectively.
Now, the field equations (\ref{4_eqn}) incorporating the FLRW spacetime (\ref{4}) with the perfect matter distribution, read as
\begin{eqnarray} \label{15}
6\frac{\dot{a}^2}{a^2}-24\frac{\dot{a}^3}{a^3}\dot{{f_\mathcal{G}}}+\mathcal{G}f_\mathcal{G}-f-2(\rho+p)f_\textit{T}=2\kappa^2\rho,\\\label{16}
-2\bigg(2\frac{\ddot{a}}{a}+\frac{\dot{a}^2}{a^2}\bigg)+16\frac{\dot{a}\ddot{a}}{a^2}\dot{{f_\mathcal{G}}}+
8\frac{\dot{a}^2}{a^2}\ddot{{f_\mathcal{G}}}-\mathcal{G}f_\mathcal{G}+f=2\kappa^2 p.
\end{eqnarray}
To keep the things simple, we assume that $f\equiv f(\mathcal{G},\mathrm{\textit{T}})$,
$f_{\mathcal{G}}\equiv f_{\mathcal{G}}(\mathcal{G},\mathrm{\textit{T}})$ etc.
Due to highly non-linearity, occasionally it becomes too hard to opt for particular $f(\mathcal{G},\mathrm{\textit{T}})$ model that may exhibit some favorable outcome either analytically or by adopting some numerical techniques. The easiest way is go for a linear combination of the form as
\begin{equation}\label{VM}
f(\mathcal{G},\mathrm{\textit{T}})=f_{1}(\mathcal{G})+f_{2}(\mathrm{\textit{T}}).
\end{equation}
The alternative emerging opportunities such as the rational expressions associated to the variables or product structures may be found as well. However, the desired outcome may not be acquired so frequently due to the involvement of complex and lengthy non-linear equations. Therefore, for our work here, we are indebted to investigate cosmological solutions through the employment of the linear form (\ref{VM}).

Here, we consider $f_{1}(\mathcal{G})=\mathcal{G}+\lambda \mathcal{G}^2$, with $\lambda$ is a real parameter. Such choice is crucial as the power law $f(\mathcal{G})$ models have been presented with viable results \cite{17}. Above all, by taking into account the study by Elizalde et al. \cite{Elizalde}, we apply $f_{2}(\mathrm{\textit{T}})=2\beta Log(\mathrm{\textit{T}})$, with $\beta$ being an arbitrary parameter. Under all these assumptions, the $f(\mathcal{G},\mathrm{\textit{T}})$ model acquires the form as follows
\begin{equation}
f(\mathcal{G},\mathrm{\textit{T}})=\mathcal{G}+\lambda \mathcal{G}^2+2\beta Log(\mathrm{\textit{T}}).
\end{equation}

\subsection{A Bird Eye View of Cosmological Parameters}
The rate of the expansion of the universe can be expressed with the help of Hubble parameter $H$ indulging the scale factor $a$, such that $H=\dot{a}/a$, with $\dot{a}$ being its first time derivative. It is known that $q$ stands for the deceleration parameter, connected through the second time derivative of scale factor $a$, the so-called state finder parameter``jerk'' is represented by $j$ and is related via the third time derivative of $a$, and the letter $s$ stands for the so-called ``snap'' parameter, relating through the fourth time derivative of $a$. These parametric quantities are crucially important for the observational cosmology, and are defined as
\begin{equation}\label{MUSH1}
q=-\frac{1}{H^2}\frac{\ddot{a}}{a};~~~~~~j=\frac{1}{H^3}\frac{\dddot{a}}{a};~~~~~~s=\frac{1}{H^4}\frac{\ddddot{a}}{a}.
\end{equation}

The Taylor expansion of the cosmological EoS around the current epoch is the most straightforward and the simplest model that may be considered as it does not assume any theoretical constraints on the true nature of the cosmological fluid. Common cosmological models sensibly based on the assumptions that once some hypothetical EoS is proposed, the Friedmann equations are implemented to establish the evolution of  FRW scale factor $a(t)$. On contrary, a postdictive approach might practically yield observational data relating to the scale factor and apply the Friedmann equations to deduce a pragmatic cosmological EoS. Particularly, the value of the scale factor and its time derivatives  determined at the current epoch put restrictions on the value along with its time  derivatives of the cosmological EoS at the recent epoch. Working out the first three EoS Taylor coefficients  at the current epoch demands an amount of the deceleration, jerk, and snap--the second, third, and the fourth time derivatives of the scale factor $a(t)$. Higher-order Taylor coefficients in the EoS are linked to the high order time derivatives of the scale factor. Since the jerk and the snap parameters are somewhat problematic to be measured due  to the involvement of the third and fourth terms in the Hubble law Taylor series expansion, it is quite evident why direct observational limitations on the cosmological EoS are so weak relatively, and are expected to remain weak for the predictable future.\\\

The deceleration, jerk and the snap parameters are dimension free quantities, and  expanding the scale factor $a$ through the Taylor series around $t_0$ gives
\begin{equation}\label{MUSH2}
a(t)=a_{0}\bigg[1+H_0(t-t_0)-\frac{1}{2}q_0H_{0}^2(t-t_0)^2+\frac{1}{6}j_0H_{0}^3(t-t_0)^3+\frac{1}{24}s_0H_{0}^4(t-t_0)^4+O[t^5]\bigg].
\end{equation}
Therefore, the luminosity distance is determined as \cite{Visser2004}.
\begin{equation}\label{MUSH3}
DL=\frac{c}{H_0}\bigg[z+\frac{1}{2}(1-q_0)z^2-\frac{1}{3!}z^3(1+j_0-q_0-3q_0^2)+\frac{1}{4!}z^4(2+s_0+5j_0+10q_0j_0-2q_0-15q_0^2-15q_0^3)+O(z^5)\bigg].
\end{equation}
For the sake of determining redshift range of SNe Ia, the terms in Eq.\ref{MUSH3} above the cubic power can be ignored. If the models under study possess the same values of the deceleration and jerk parameters, one may have the degeneracy of these models from Eq.\ref{MUSH3}. Therefore,  the snap parameters must be measured to differentiate between the models. This ultimately requires the high redshift
objects. Our main interest lies in the four statefinder quantities, namely the Hubble parameter $H$, the deceleration
parameter $q$, the snap parameter $s$,  and the jerk parameter $j$. The connections among these parametric quantities  q(z), j(z) and s(z) are the functions of the Hubble parameter as given below

\begin{equation}\label{MUSH4}
q=-\frac{\dot{H}}{H^2}-1,
\end{equation}
\begin{equation}\label{MUSH5}
j=-2-3q+\frac{\ddot{H}}{H^3},
\end{equation}
\begin{equation}\label{MUSH6}
s=\frac{j-1}{3(q-\frac{1}{2})}.
\end{equation}
The last two, namely the jerk $j$ and snap $s$ possess the simple numerical values for the $\Lambda{CDM}$ model, that is $s=0$, and $j=1$. As for the observational cosmology has been concerned, we still have no direct access to the scale factor $a(t)$ through the entire history of  our universe.  However, we have imprecise access to the present values of the scale factor and its derivatives, as determined in the Hubble parameter, the deceleration parameter, the snap parameter,  and the jerk parameter. This more inadequate information may still be applied to find out some useful information about the  EoS.
\section{Exact Solutions}
\begin{figure}\center
\begin{tabular}{cccc}
\epsfig{file=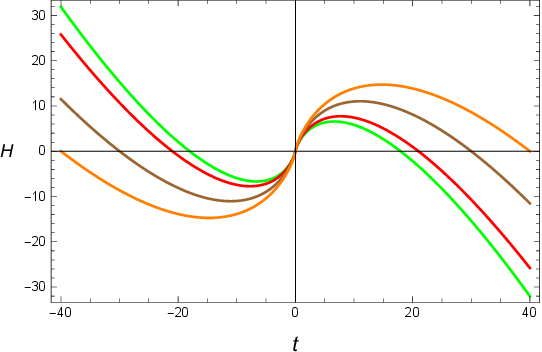,width=0.38\linewidth} \\
\end{tabular}
\caption{Evolution of Hubble Parameter with $\alpha =1$; $\gamma =-0.0005$;~ $c=18(\textcolor{green}{\bigstar})$; $c=21(\textcolor{red}{\bigstar})$; $c=30(\textcolor{brown}{\bigstar})$; $c=40(\textcolor{orange}{\bigstar})$.}\center
\label{Fig:51}
\end{figure}
To find some viable cosmological solutions, we choose a well-known form of the Hubble parameter \cite{Bouncing7}
\begin{equation}\label{HP1}
H (t)=\alpha ~t~h(t).
\end{equation}
Here $h(t)$ is an arbitrary smooth function. Fuhrer, we assume a particular form of $h(t)$ as follows:
\begin{equation}\label{HP2}
h (t)=Log \left|\frac{c-\gamma  \cot ^{-1}(t)}{t}\right|,
\end{equation}
where $\alpha$, $c$ and $\gamma$ are arbitrary real parameters. The complete parameterized form of the Hubble parameter turns out to be
\begin{equation}\label{HP3}
H(t)=\alpha ~ t ~Log \left|\frac{c-\gamma  \cot ^{-1}(t)}{t}\right|.
\end{equation}
\begin{figure}\center
\begin{tabular}{cccc}
\epsfig{file=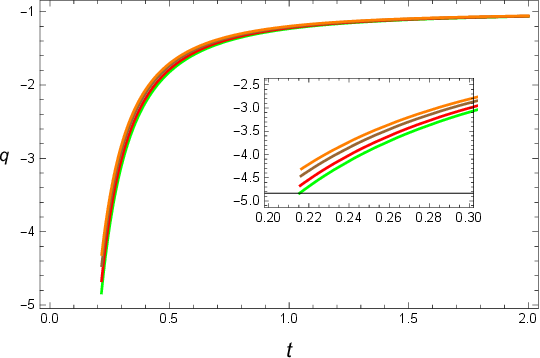,width=0.38\linewidth}
\end{tabular}
\caption{Evolution of Deceleration Parameter with $\alpha =1$; $\gamma =-0.0005$;~ $c=18(\textcolor{green}{\bigstar})$; $c=21(\textcolor{red}{\bigstar})$; $c=30(\textcolor{brown}{\bigstar})$; $c=40(\textcolor{orange}{\bigstar})$.}\center
\label{Fig:51}
\end{figure}
The parameter $\alpha$ has two purposes, either it can be used for scaling or for the change of phase. Moreover, parameter $c$ is also used for scaling along time axis.
This form of Hubble parameter is almost similar as already proposed in literature to obtain some interesting results while studying $f(\mathcal{R},\mathrm{\textit{T}})$ theory of gravity \cite{Bouncing7}. The evolution of Hubble parameter is shown in Fig. \textbf{1} for positive $\alpha$ and some different values of $c$. We can see real bounce behavior from contraction to expansion phase.
Using Eqs.(\ref{HP3}) and (\ref{MUSH4}), the mathematical expression for deceleration parameter takes the form
\begin{equation}
q =-\frac{\frac{\gamma  t}{\left(t^2+1\right) \left(c-\gamma  \cot ^{-1}(t)\right)}+Log \left|\frac{c-\gamma  \cot ^{-1}(t)}{t}\right|-1}{\alpha  t^2 Log ^2\left|\frac{c-\gamma  \cot ^{-1}(t)}{t}\right|}-1
\end{equation}
\begin{figure}\center
\begin{tabular}{cccc}
\epsfig{file=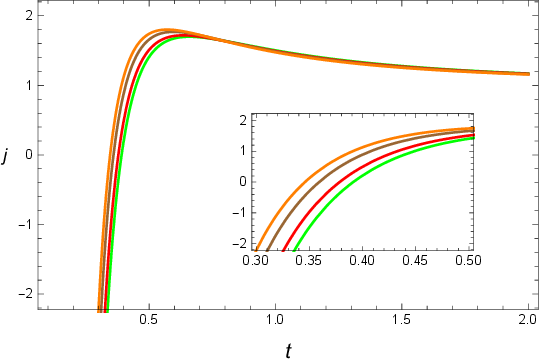,width=0.38\linewidth}&
\epsfig{file=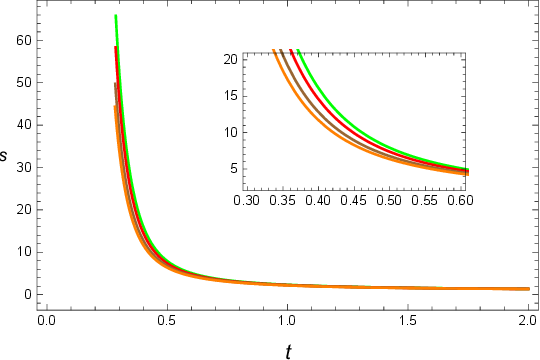,width=0.38\linewidth} \\
\end{tabular}
\caption{Evolution of Jerk and Snap Parameters with $\alpha =1$; $\gamma =-0.0005$;~ $c=18(\textcolor{green}{\bigstar})$; $c=21(\textcolor{red}{\bigstar})$; $c=30(\textcolor{brown}{\bigstar})$; $c=40(\textcolor{orange}{\bigstar})$.}\center
\label{Fig:51}
\end{figure}
The negative behavior of deceleration parameter can be seen in Fig. \textbf{2}. It is clear that as the time grows, $q\rightarrow -1$, which shows that the results are in good agrement with recent observed value ($q_0 = -0.81\pm0.14$).
Similarly, using Eqs.(\ref{HP3}) and (\ref{MUSH4}), the mathematical expression for jerk parameter becomes
\begin{eqnarray} \nonumber
j=\frac{2 \left(c^2 \left(t^2+1\right)^2-2 c \gamma  t^3+\gamma  \cot ^{-1}(t) \left(-2 c \left(t^2+1\right)^2+2 \gamma  t^3+\gamma  \left(t^2+1\right)^2 \cot ^{-1}(t)\right)-\gamma ^2 t^2\right)}{\alpha ^2 t^4 \left(t^2+1\right)^2 \left(c-\gamma  \cot ^{-1}(t)\right)^2 {Log} ^3\left|\frac{c-\gamma  \cot ^{-1}(t)}{t}\right|}+\frac{3}{\alpha  t^2 {Log} \left(\frac{c-\gamma  \cot ^{-1}(t)}{t}\right)}-\\\nonumber\frac{2 \left(-c \left(t^2+1\right)+\gamma  \left(t^2+1\right) \cot ^{-1}(t)+\gamma  t\right)^2}{\alpha ^2 t^4 \left(t^2+1\right)^2 \left(c-\gamma  \cot ^{-1}(t)\right)^2 {Log} ^4\left|\frac{c-\gamma  \cot ^{-1}(t)}{t}\right|}+\frac{-c \left(t^2+1\right) \left(3 \alpha  t^2+2\right)+3 \alpha  \gamma  t^3+\gamma  \left(t^2+1\right) \left(3 \alpha  t^2+2\right) \cot ^{-1}(t)}{\alpha ^2 t^4 \left(t^2+1\right) \left(c-\gamma  \cot ^{-1}(t)\right) {Log} ^2\left|\frac{c-\gamma  \cot ^{-1}(t)}{t}\right|}+1,\\
\end{eqnarray}
The expression for snap parameter is too much lengthy to be presented here. The evolution of both jerk and snap parameters is reflected in Fig. \textbf{3}.
\begin{figure}\center
\begin{tabular}{cccc}
\epsfig{file=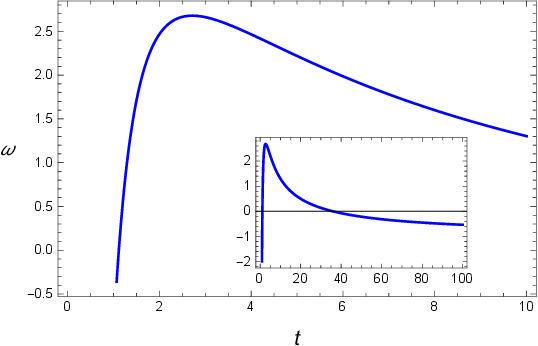,width=0.38\linewidth} &
\epsfig{file=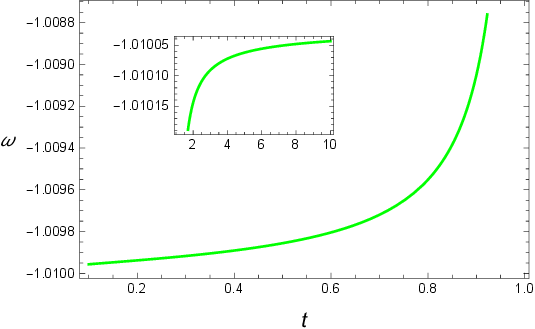,width=0.38\linewidth} \\
\end{tabular}
\caption{Evolution of EoS Parameters: Left plot for  $\omega_1 (t)=-\frac{k~Log~(t+\epsilon)}{t}-1$, $\epsilon =0.000001$; $k = -10$; Right plot for $\omega_2 (t)=\frac{r}{Log ~t}-s$, $r=-0.0001,~s=1.01$}\center
\label{Fig:51}
\end{figure}
EoS parameter is considered to be one of the important pillars on which cosmological dynamics stand. In particular, the role EoS parameter becomes more crucial while studying the modified theories of gravity due to the complicated nature of field equations. For example, in our present study, the field equations become highly nonlinear and difficult to handle due to the involvement of $\lambda \mathcal{G}^2$ and logarithmic trace terms. It is worthwhile to mention here that the explicit expression for energy density and pressure profiles are not obtained through analytical methods. Thus the only choice left with us is to consider an appropriate form of EoS parameter for further analysis. Following the interesting proposals by Elizalde et al. \cite{Elizalde}, we firstly consider the possibility of some bouncing solutions with the EoS parameter as follows:
\begin{equation}\label{EoS1}
\omega_1 (t)=-\frac{k~Log~(t+\epsilon)}{t}-1,
\end{equation}
where $\epsilon$ is very small real parameter and $k$ is an arbitrary constant. It is worthwhile to mention here that this EoS parameter is the usual one defined for the matter only, i.e. the ratio of pressure and energy density.
Further, it is to be noticed that $\omega_1$ extends
from negative infinity as $t \rightarrow 0$ to $\omega_1= -1$ (cosmic expansion phase) when $t = 1 -\epsilon$.
Our second choice for EoS parameter is
\begin{equation}\label{EoS2}
\omega_2 (t)=\frac{r}{Log ~t}-s,
\end{equation}
where $r$ is a negative while $s$ is a positive parameter. Here, we can also see that
$\omega_2$ varies from negative infinity as $t \rightarrow 1$  to the cosmic expansion era at $t = e^{\frac{r}{s-1}}$
and moves on, eventually coming back to again the same phase as $t$ approaches positive infinity and $s = 1$. The behavior of these EoS parameters can be seen in Fig. \textbf{4}. Elizalde et al. \cite{Elizalde} obtained viable cosmological solutions in the framework of \textbf{$f(R,T)$} theory of gravity using these two interesting choices of EoS parameter. In this study, we extend their work in the context of $f(\mathcal{G},\mathrm{\textit{T}})$  gravity and in particular, by considering the Hubble parameter (\ref{HP3}). Now, we discuss bouncing solutions using these two different EoS parameters.

\subsection{Bouncing Solutions with EoS $\omega=-\frac{k~Log~(t+\epsilon)}{t}-1$}

In this subsection, we present the evolution of physical parameters.
Using EoS (\ref{EoS1}), the expressions for energy density and pressure profiles (\ref{15})-(\ref{16}) take the forms
\begin{eqnarray}\nonumber
\rho &=&-\frac{2t}{k Log (t+\epsilon ) (3 k Log (t+\epsilon )+4 t)}\times \big(288 k \lambda  \dddot{H} H^4 Log (t+\epsilon )+384 \lambda  t \dddot{H} H^4+864 k \lambda  H^5 \ddot{H} Log (t+\epsilon )+\\\nonumber&&1152 \lambda  t H^5 \ddot{H}-1152 k \lambda  H^6 \dot{H} Log (t+\epsilon )+5184 k \lambda  H^4 \dot{H}^2 Log (t+\epsilon )+1728 k \lambda  H^2 \dot{H}^3 Log (t+\epsilon )-\\\nonumber&&3 k \dot{H} Log (t+\epsilon )-1536 \lambda  t H^6 \dot{H}+6912 \lambda  t H^4 \dot{H}^2+2304 \lambda  t H^2 \dot{H}^3-4 t \dot{H}+2304 k \lambda  H^3 \dot{H} \ddot{H} Log (t+\epsilon )+\\\label{fe1}&&3072 \lambda  t H^3 \dot{H} \ddot{H}+\beta  k Log (t+\epsilon )\big),
\end{eqnarray}
\begin{eqnarray}\nonumber
p&=&\frac{2 (k Log (t+\epsilon )+t)}{k Log (t+\epsilon ) (3 k Log (t+\epsilon )+4 t)}\times\big(288 k \lambda  \dddot{H} H^4 Log (t+\epsilon )+384 \lambda  t \dddot{H} H^4+864 k \lambda  H^5 \ddot{H} Log (t+\epsilon )+\\\nonumber&&1152 \lambda  t H^5 \ddot{H}-1152 k \lambda  H^6 \dot{H} Log (t+\epsilon )+5184 k \lambda  H^4 \dot{H}^2 Log (t+\epsilon )+1728 k \lambda  H^2 \dot{H}^3 Log (t+\epsilon )-\\\nonumber&&3 k \dot{H} Log (t+\epsilon )-1536 \lambda  t H^6 \dot{H}+6912 \lambda  t H^4 \dot{H}^2+2304 \lambda  t H^2 \dot{H}^3-4 t \dot{H}+2304 k \lambda  H^3 \dot{H} \ddot{H} Log (t+\epsilon )+\\\label{fe2}&&3072 \lambda  t H^3 \dot{H} \ddot{H}+\beta  k Log (t+\epsilon )\big),
\end{eqnarray}
The graphical behavior of energy density and pressure using Eqs. (\ref{fe1},\ref{fe2}) and Hubble parameter (\ref{HP3}) is
is shown in Fig. \textbf{5}. We have assumed same values of parameters which are used in the evolutions of above mentioned cosmological parameters (for details see the caption under the Fig. \textbf{5}).
Left plot of Fig. \textbf{5} shows that the energy density is positive within the neighborhood of bouncing point $t = 0$ while the right graph suggests that the pressure profile is negative, which justifies the current cosmic expansion.
\begin{figure}\center
\begin{tabular}{cccc}
\epsfig{file=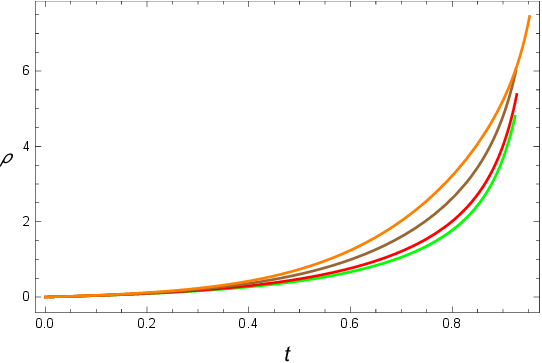,width=0.38\linewidth} &
\epsfig{file=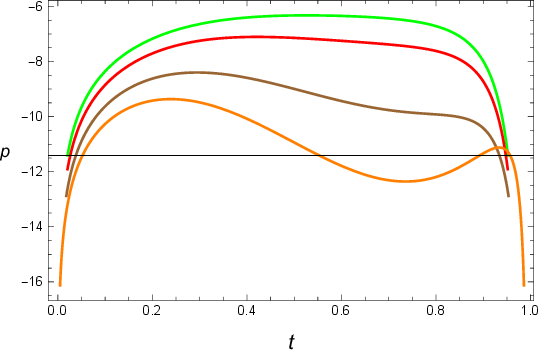,width=0.38\linewidth} \\
\end{tabular}
\caption{Energy density and pressure profiles for EoS $\omega_1 (t)=-\frac{k~Log~(t+\epsilon)}{t}-1$ with $\alpha =1$; $\epsilon =0.000001$; $\gamma =-0.0005;~ k = -10;~\beta =0.0000005;~\lambda =-0.000005$; $c=18(\textcolor{green}{\bigstar})$; $c=21(\textcolor{red}{\bigstar})$; $c=30(\textcolor{brown}{\bigstar})$; $c=40(\textcolor{orange}{\bigstar})$.}\center
\label{Fig:51}
\end{figure}
\begin{figure}\center
\begin{tabular}{cccc}
\epsfig{file=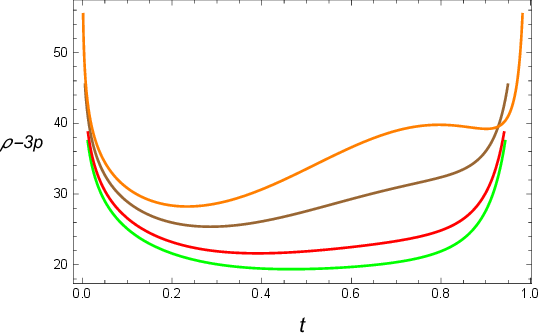,width=0.38\linewidth} &
\epsfig{file=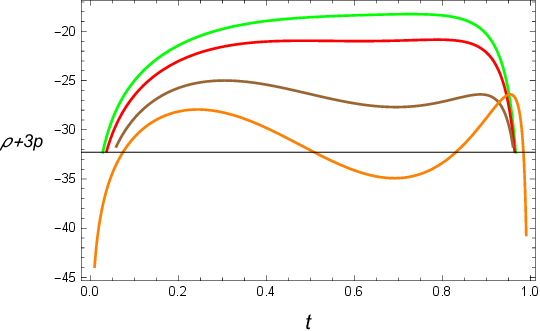,width=0.38\linewidth} \\
\end{tabular}
\caption{Validation of TEC and SEC for EoS $\omega_1 (t)=-\frac{k~Log~(t+\epsilon)}{t}-1$ with $\alpha =1$; $\epsilon =0.000001$; $\gamma =-0.0005;~ k = -10;~\beta =0.0000005;~\lambda =-0.000005$; $c=18(\textcolor{green}{\bigstar})$; $c=21(\textcolor{red}{\bigstar})$; $c=30(\textcolor{brown}{\bigstar})$; $c=40(\textcolor{orange}{\bigstar})$.}\center
\label{Fig:51}
\end{figure}
It is important to mention here that the choice of model parameters strictly depends upon the evolution of cosmological parameters and in particular, the conservation equation. The trace energy condition (TEC) was popular among the researchers during the decade of $1960$. The study of this energy condition was not too much cited in the recent literature. TEC is seen to be satisfied while SEC is violated as shown in Fig. \textbf{6} for $f(\mathcal{G},\mathrm{\textit{T}})$ gravity model under discussion.
\begin{figure}\center
\begin{tabular}{cccc}
\epsfig{file=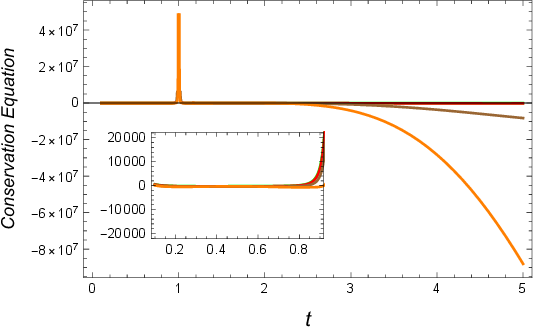,width=0.38\linewidth} \\
\end{tabular}
\caption{Validation of Conservation Equation for EoS $\omega_1 (t)=-\frac{k~Log~(t+\epsilon)}{t}-1$ with $\alpha =1$; $\epsilon =0.000001$; $\gamma =-0.0005;~ k = -10;~\beta =0.0000005;~\lambda =-0.000005$; $c=18(\textcolor{green}{\bigstar})$; $c=21(\textcolor{red}{\bigstar})$; $c=30(\textcolor{brown}{\bigstar})$; $c=40(\textcolor{orange}{\bigstar})$.}\center
\label{Fig:51}
\end{figure}

Modified theories involving matter curvature coupling do obey the usual conservation equation of GR. This issue in case of $f(\mathcal{G},\mathrm{\textit{T}})$ theory is evident by Eq. (\ref{div}). In fact, this seems to be a drawback that the theory might be plagued by divergences at astrophysical scales. One of the important contributions of the present study is the investigation of conservation equation in the framework of parameters used in the solutions. Following condition on Eq.(\ref{div}) can be imposed to deal with the issue.
\begin{equation}\label{gaf}
\nabla^{\xi}\Theta_{\xi\eta}-
\frac{g_{\xi\eta}}{2}\nabla^{\xi}\textit{T}+
(\mathrm{\textit{T}}_{\xi\eta}+\Theta_{\xi\eta})
\nabla^{\xi}(\text{Log}f_{\mathrm{\textit{T}}}(\mathcal{G},\mathrm{T}))=0.
\end{equation}
\begin{figure}\center
\begin{tabular}{cccc}
\epsfig{file=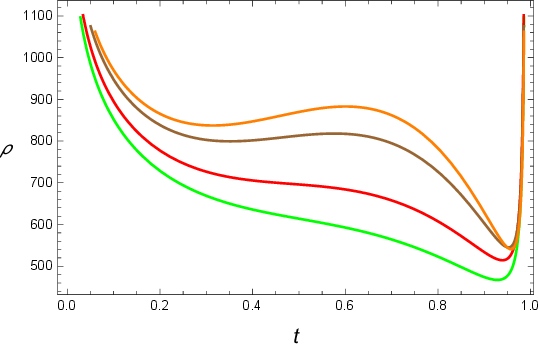,width=0.38\linewidth} &
\epsfig{file=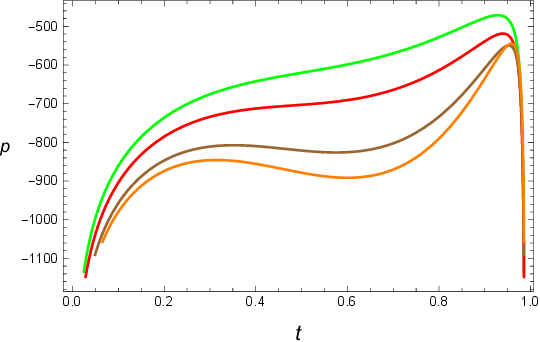,width=0.38\linewidth} \\
\end{tabular}
\caption{Energy density and pressure profiles for EoS $\omega_2 (t)==\frac{r}{Log ~t}-s$ with $\alpha =1$; $s =1.01$; $\gamma =-0.0005;~ r = -0.001;~\beta =0.0000005;~\lambda =-0.000005$; $c=18(\textcolor{green}{\bigstar})$; $c=21(\textcolor{red}{\bigstar})$; $c=30(\textcolor{brown}{\bigstar})$; $c=40(\textcolor{orange}{\bigstar})$.}\center
\end{figure}
We have tried to obtain exact analytic solutions of this equation, however, we do not succeed due to highly non-linear terms. So, in the present study, we have set the  model parameters in such a way that Eq. (\ref{gaf}) seems to be satisfied partially as shown in Fig. \textbf{7}. A magnified view is also inserted in the figure for a clear understanding. It is interesting to notice that conservation equation is satisfied in the neighborhood of initial bouncing point but deviates as the time grows. In fact, the deviations are more for larger values of $c$ as compared to smaller values.

\subsection{Bouncing Solutions with EoS $\omega=\frac{r}{Log ~t}-s$}

Here, using EoS (\ref{EoS2}), the expressions for energy density and pressure profiles (\ref{15})-(\ref{16}) take the forms
\begin{eqnarray}\nonumber
\rho &=&-\frac{1}{{8 \beta ^2+6 \beta +1}}\times\big(-192 \beta  \lambda  \dddot{H} H^4+576 \beta  \lambda  H^5 \ddot{H}+576 \lambda  H^5 \ddot{H}+4224 \beta  \lambda  H^6 \dot{H}-1728 \beta  \lambda  H^4 \dot{H}^2-1152 \beta  \lambda  H^2 \dot{H}^3\\&&+2 \beta  \dot{H}+1728 \lambda  H^6 \dot{H}+864 \lambda  H^4 \dot{H}^2-1536 \beta  \lambda  H^3 \dot{H} \ddot{H}-576 \beta  \lambda  H^8-6 \beta  H^2-288 \lambda  H^8-3 H^2\big),\label{fe1a}
\end{eqnarray}
\begin{eqnarray}\nonumber
p&=&-\frac{1}{8 \beta ^2+6 \beta +1}\big(-576 \beta  \lambda  \dddot{H} H^4-192 \lambda  \dddot{H} H^4-2880 \beta  \lambda  H^5 \ddot{H}-1152 \lambda  H^5 \ddot{H}-1152 \beta  \lambda  H^6 \dot{H}-12096 \beta  \lambda  H^4 \dot{H}^2-\\\nonumber&&3456 \beta  \lambda  H^2 \dot{H}^3+6 \beta  \dot{H}-960 \lambda  H^6 \dot{H}-4320 \lambda  H^4 \dot{H}^2-1152 \lambda  H^2 \dot{H}^3+2 \dot{H}-4608 \beta  \lambda  H^3 \dot{H} \ddot{H}-1536 \lambda  H^3 \dot{H} \ddot{H}+\\&&576 \beta  \lambda  H^8+6 \beta  H^2+288 \lambda  H^8+3 H^2\big),\label{fe1b}
\end{eqnarray}
The graphical behavior of energy density and pressure using Eqs. (\ref{fe1a},\ref{fe1b}) and Hubble parameter (\ref{HP3})
is shown in Fig. \textbf{8}. Left graph of Fig. \textbf{8} depicts that the energy density is positive within the neighborhood of bouncing point $t = 0$ while the right plot indicates that the pressure profile is negative. It is shown in the left plot of Fig. \textbf{9} that TEC is satisfied all the time. However, SEC is not obeyed as shown in the right plot of Fig. \textbf{9}.
\begin{figure}\center
\begin{tabular}{cccc}
\epsfig{file=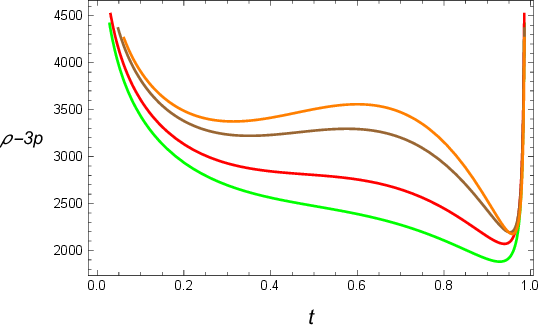,width=0.38\linewidth} &
\epsfig{file=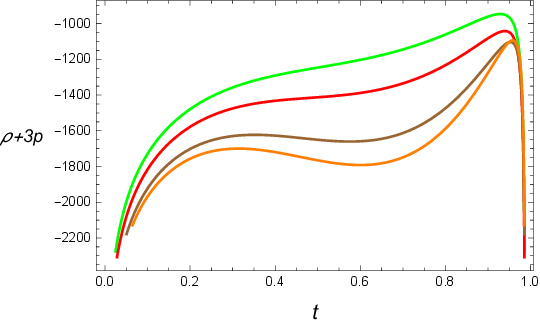,width=0.38\linewidth} \\
\end{tabular}
\caption{Validation of TEC and SEC for EoS $\omega_2 (t)==\frac{r}{Log ~t}-s$ with $\alpha =1$; $s =1.01$; $\gamma =-0.0005;~ r = -0.001;~\beta =0.0000005;~\lambda =-0.000005$; $c=18(\textcolor{green}{\bigstar})$; $c=21(\textcolor{red}{\bigstar})$; $c=30(\textcolor{brown}{\bigstar})$; $c=40(\textcolor{orange}{\bigstar})$.}\center
\label{Fig:51}
\end{figure}
\begin{figure}\center
\begin{tabular}{cccc}
\epsfig{file=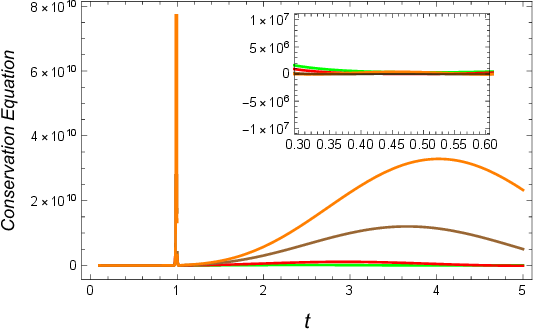,width=0.38\linewidth} &
\end{tabular}
\caption{Validation of Conservation Equation for EoS $\omega_2 (t)==\frac{r}{Log ~t}-s$ with $\alpha =1$; $s =1.01$; $\gamma =-0.0005;~ r = -0.001;~\beta =0.0000005;~\lambda =-0.000005$; $c=18(\textcolor{green}{\bigstar})$; $c=21(\textcolor{red}{\bigstar})$; $c=30(\textcolor{brown}{\bigstar})$; $c=40(\textcolor{orange}{\bigstar})$.}\center
\label{Fig:51}
\end{figure}
Finally, Fig. \textbf{10} shows that the conservation equation is satisfied in the neighborhood of initial bouncing point. It is important to mention here that this type of partially satisfied behavior is strictly dependent on the choice of model parameters. The more we do change the parameters, the more bad behavior of conservation equation may be obtained. It can be seen that conservation is satisfied for all values of $c$ after the initial bounce, till the singularity due to chosen EoS model (a magnified view is also inserted in the figure for better understanding). After that some deviations may be observed later on for the larger values of $c$. All the above discussions and graphical analysis for both cases indicate that our proposed $f(\mathcal{G},\mathrm{\textit{T}})$ gravity models provide some good bouncing solutions with the chosen EoS and model parameters.

\section{Concluding Remarks}

In recent few years, the Gauss-Bonnet $f(\mathcal{G},\mathrm{\textit{T}})$ theory of gravity has impressed many researchers owing to its coupling of trace of the stress-energy tensor with the Gauss-Bonnet term. In this context, we focuss ourselves to study bouncing universe with in $f(\mathcal{G},\mathrm{\textit{T}})$ gravity background. In particular, we choose Gauss-Bonnet cosmological model with logarithmic trace term, i.e.
$f(\mathcal{G},\mathrm{\textit{T}})=\mathcal{G}+\lambda \mathcal{G}^2+2\beta Log(\mathrm{\textit{T}})$ and explore the possibility of some bouncing Universe by considering a well known from of Hubble parameter \cite{Bouncing7}.
The discussion of some important cosmological parameters is included to develop the results. The exact bouncing solutions with physical analysis are studied with the choice of two equation of state parameters. The main results of current study are listed below:\\\\
\begin{itemize}
\item This study is based upon the Hubble parameter $H(t)=\alpha ~ t ~Log \left|\frac{c-\gamma  \cot ^{-1}(t)}{t}\right| $, where the real parameter $\alpha$ has two roles, firstly it may be used for the scaling purposeless. The second use is for the change of phase.
The bouncing cosmology is studied using a Gauss-Bonnet cosmological model with logarithmic trace term. The evolution of Hubble, deceleration, jerk and snap parameters is provided in Figs. \textbf{1-3}.
The negative behavior of deceleration parameter is witnessed and it is clear that as the time grows, $q\rightarrow -1$, which shows that the results are in good agrement with recent observed value ($q_0 = -0.81\pm0.14$). Similarly, the evolution of both jerk and snap parameters is reflected in Fig. \textbf{3}.
\item
Further, the two EoS parameters $\omega_1 (t)=-\frac{k~Log~(t+\epsilon)}{t}-1$ and $ \omega_2 (t)=\frac{r}{Log ~t}-s$ are chosen. It is worthwhile to notice that EoS parameter extends from negative infinity as $t \rightarrow 0$ to $\omega_1= -1$ (cosmic expansion phase) when $t = 1 -\epsilon$.
As far as the second choice for EoS parameter is concerned, we can also see that
$\omega_2$ varies from negative infinity as $t \rightarrow 1$  to the cosmic expansion era at $t = e^{\frac{r}{s-1}}$
and moves on, eventually coming back to again the same phase as $t$ approaches positive infinity and $s = 1$. The behavior of these EoS parameters can be seen in Fig. \textbf{4}.
\item
The graphical behavior of energy density and pressure using Eqs. (\ref{fe1},\ref{fe2}) and Hubble parameter (\ref{HP3}) is
is shown in Fig. \textbf{5}. We have assumed same values of parameters which are used in the evolutions of above mentioned cosmological parameters (for details see the caption under the Fig. \textbf{5}).
Left plot of Fig. \textbf{5} shows that the energy density is positive within the neighborhood of bouncing point $t = 0$ while the right graph suggests that the pressure profile is negative, which justifies the current cosmic expansion.
It is important to mention here that the choice of model parameters strictly depends upon the evolution of cosmological parameters and in particular, the conservation equation. TEC is seen to be satisfied while SEC is violated as shown in Fig. \textbf{6} for $f(\mathcal{G},\mathrm{\textit{T}})$ gravity model under discussion.
Moreover, the graphical behavior of energy density and pressure using Eqs. (\ref{fe1a},\ref{fe1b}) and Hubble parameter (\ref{HP3}) is
is shown in Fig. \textbf{8}. Left graph of Fig. \textbf{8} depicts that the energy density is positive within the neighborhood of bouncing point $t = 0$ while the right plot indicates that the pressure profile is negative. It is shown in the left plot of Fig. \textbf{9} that TEC is satisfied all the time. However, SEC is not obeyed as shown in the right plot of Fig. \textbf{9}.

\item
Modified theories involving matter curvature coupling do obey the usual conservation equation of GR. This issue in case of $f(\mathcal{G},\mathrm{\textit{T}})$ theory is evident by Eq. (\ref{div}). In fact, this seems to be a drawback that the theory might be plagued by divergences at astrophysical scales. One of the important contributions of the present study is the investigation of conservation equation in the framework of parameters used in the solutions. Following condition on Eq.(\ref{div}) can be imposed to deal with the issue.
\begin{equation}\label{gaf}
\nabla^{\xi}\Theta_{\xi\eta}-
\frac{g_{\xi\eta}}{2}\nabla^{\xi}\textit{T}+
(\mathrm{\textit{T}}_{\xi\eta}+\Theta_{\xi\eta})
\nabla^{\xi}(\text{Log}f_{\mathrm{\textit{T}}}(\mathcal{G},\mathrm{T}))=0.
\end{equation}
We have tried to obtain exact analytic solutions of this equation, however, we do not succeed due to highly non-linear terms. So in the present study, we have set the  model parameters in such a way that Eq. (\ref{gaf}) seems to be satisfied partially as shown in Figs. \textbf{7} and \textbf{10}. A magnified view is also inserted in the figures for a clear understanding. It is interesting to notice that conservation equation is satisfied in the neighborhood of initial bouncing point but deviates as the time grows. In fact, the deviations are more for larger values of $c$ as compared to smaller values.\\\\
\end{itemize}
All the above points discussed above show that our proposed $f(\mathcal{G},\mathrm{\textit{T}})$ gravity model provide good bouncing solutions with the chosen EoS and cosmological parameters. It is shown that the results do agree with the present values of deceleration, jerk and snap parameters. Moreover, it is concluded that the model parameters are quite important for the validity of conservation equation (as the matter coupled theories do not obey the usual conservation law).
\\\\
%

\end{document}